\documentclass[fleqn,10pt]{wlscirep}
\usepackage[utf8]{inputenc}
\usepackage[T1]{fontenc}
\usepackage{graphicx}
\usepackage{dcolumn}
\usepackage{bm}
\usepackage{amsmath}
\usepackage{graphicx}
\usepackage{subfigure}
\usepackage{multirow}
\usepackage{xcolor}

\newcommand{\change}[1]{\textcolor{black}{{#1}}}  
\newcommand{\changeII}[1]{\textcolor{black}{{#1}}}  

\title{Local dominance unveils clusters in networks}

\author[1$\dagger$]{Dingyi Shi} 
\author[1$\dagger$]{Fan Shang} 
\author[1,2,3*]{Bingsheng Chen}
\author[4]{Paul Expert} 
\author[5,6*]{Linyuan L\"u}
\author[7]{H. Eugene Stanley} 
\author[8,9]{Renaud Lambiotte} 
\author[2]{Tim S.\ Evans} 
\author[1*]{Ruiqi Li}

\affil[1]{UrbanNet Lab, College of Information Science and Technology, Beijing University of Chemical Technology, Beijing 100029, China}
\affil[2]{centre for Complexity Science, Imperial College London, London SW7 2AZ, UK}
\affil[3]{Network Science Institute and Department of Physics, Northeastern University, Boston, MA 02115, USA}
\affil[4]{Global Business School for Health, University College London, London WC1E 6BT, UK}
\affil[5]{School of Cyberspace Security, University of Science and Technology of China, Hefei 230026, China}
\affil[6]{Institute of Fundamental and Frontier Sciences, University of Electronic Science and Technology of China, Chengdu 610054, China}
\affil[7]{Center for Polymer Studies and Physics Department, Boston University, Boston, MA 02215, USA}
\affil[8]{Mathematical Institute, University of Oxford, Oxford OX2 6GG, UK}
\affil[9]{Turing Institute, London NW1 2DB, UK}

\affil[$\dagger$]{F.S. and D.S. contributed equally to this work.}
\affil[*]{corresponding authors: lir@buct.edu.cn (Ruiqi Li), linyuan.lv@ustc.edu.cn (Linyuan L\"u), bi.chen@northeastern.edu (Bingsheng Chen).}

\begin{abstract}
Clusters or communities can provide a coarse-grained description of complex systems at multiple scales, but their detection remains challenging in practice. Community detection methods often define communities as dense subgraphs, or subgraphs with few connections in-between, via concepts such as the cut, conductance, or modularity. Here we consider another perspective built on the notion of local dominance, where low-degree nodes are assigned to the basin of influence of high-degree nodes, and design an efficient algorithm based on local information. Local dominance gives rises to community centers, and uncovers local hierarchies in the network. Community centers have a larger degree than their neighbors and are sufficiently distant from other centers. The strength of our framework is demonstrated on synthesized and empirical networks with ground-truth community labels. The notion of local dominance and the associated asymmetric relations between nodes are not restricted to community detection, and can be utilised in clustering problems, as we illustrate on networks derived from vector data. 
\end{abstract}
\begin{document}

\flushbottom
\maketitle


\section*{Introduction} 
Many real-world datasets can be viewed as a collection of objects embedded into a global metric space, thereby providing a vector representation \cite{manning2008inforetrieval}. Alternatively, networks have become another fundamental way to model complex systems with a focus on direct pairwise interactions between constituents \cite{albert2002statistical,barabasi2016network,newman2018networks}. 
In the case of social systems, for instance, these complementary representations may correspond to a set of socio-demographic variables for each individual, e.g., in a Blau space \cite{mcpherson1983ecology}, or to a social network of interactions between individuals, e.g., via a mobile communication network \cite{blondel2015survey} or spatio-temporal co-occurrence interactions \cite{liu2022revealing}. 
In each representation, real-world systems tend to exhibit groups: regions of high density in the spatial representation, known as clusters, or high-density subgraphs in the network, known as communities. 
Such cluster or community structure provides a coarse-grained representation of the underlying complex system\cite{girvan2002community,blondel2008fast,fortunato2010community,fortunato2016community}, often associated to different functions and impacting its collective behaviours \cite{holme2003subnetwork,palla2005uncovering,clauset2008hierarchical}, and their unsupervised detection is thus essential in different areas of data science \cite{manning2008inforetrieval,fortunato2010community}.

In the vector representation, the introduction of a dissimilarity function and ideally of a distance in a metric space, provides a natural way to identify the center of a cluster, e.g., the medoid in a general metric space \cite{kaufman2009finding,rodriguez2014clustering}, and a hierarchy would form within a cluster between central and other more peripheral nodes, implying an asymmetric relationship between them. 
On the other hand, in the case of asymmetric pairwise interactions, which can be associated to an implicit hierarchy \cite{li2015social} and have long been recognized \cite{blondel2008local,serrano2009extracting,stanoev2011influence,lee2021uncovering,li2021gravity} in various network systems, community detection methods for networks place much less emphasis on the concept of community center and hierarchy within communities. 
We can always use network centrality measures on the subgraphs identified as communities to identify core and peripheral nodes \textit{a posteriori}, but these roles are not central to community detection \cite{newman2006modularity,fortunato202220}, in stark contrast to clustering methods based on embedding the data in a metric space. 

In this paper, we propose a community detection algorithm in networks, Local Search (LS), that explicitly uses the notion of local dominance and identifies community centers based on local information.  
In our method, every node is given at most one parent node deemed to be higher up in a partial ranking. 
Nodes that have a dominant position in their immediate neighborhood \cite{blondel2008local} or even beyond are identified as local leaders \cite{blondel2008local}. 
This defines a rooted tree that spans the network and gives rise to community centers that are \textit{local leaders} \cite{blondel2008local} with both a larger degree than the nodes in their basin of attraction and a relatively long distance to other local leaders higher up in the ranking. 
Our approach possesses several interesting properties. Firstly, it provides a new perspective on community detection and delivers community centers and a hierarchy within the community and even a hierarchy among communities as an explicit part of our algorithm, and so mimics advantageous features of the methods based on embedding data in a metric space. 
Secondly, the identification of communities through local dominance is highly efficient, as it uses purely local topological information and breadth-first search, and runs in linear time. The method does not require the heuristic optimization of an objective function that relies on a global null model\cite{blondel2008fast,clauset2004finding,rosvall2008maps,peel2017ground,duch2005community,mclachlan2007algorithm} or computationally costly spreading dynamics\cite{raghavan2007near,frey2007clustering,Delvenne_Yaliraki_Barahona_2010}. 
Also, our method does not rely on a similarity measure for which there is a wide choice, with an associated uncertainty and variability in results, such as is found in hierarchical clustering based methods \cite{girvan2002community,ravasz2002hierarchical,sales2007extracting,clauset2008hierarchical}. 
Finally, LS is not as susceptible to noise as most methods\cite{hoffmann2020community,fortunato2010community}, and is less therefore susceptible to finding spurious communities in random graph model realisations\cite{reichardt2006networks}.

We demonstrate the strength of LS on several classical but challenging synthetic benchmarks and on standard empirical networks with known ground-truth community labels. Our numerical evaluation also includes network representations derived from vector data.
As the LS method naturally provides community centers and local hierarchies, it creates an explicit analogy with the notion of cluster centers and distances within clusters that are found in vector clustering methods. Moreover, we also show that applying LS on discretised version of data cloud points outperforms classical unsupervised vector data clustering methods on benchmarks \cite{rodriguez2014clustering}.

\begin{figure}[!htbp]\centering
\includegraphics[width=\linewidth]{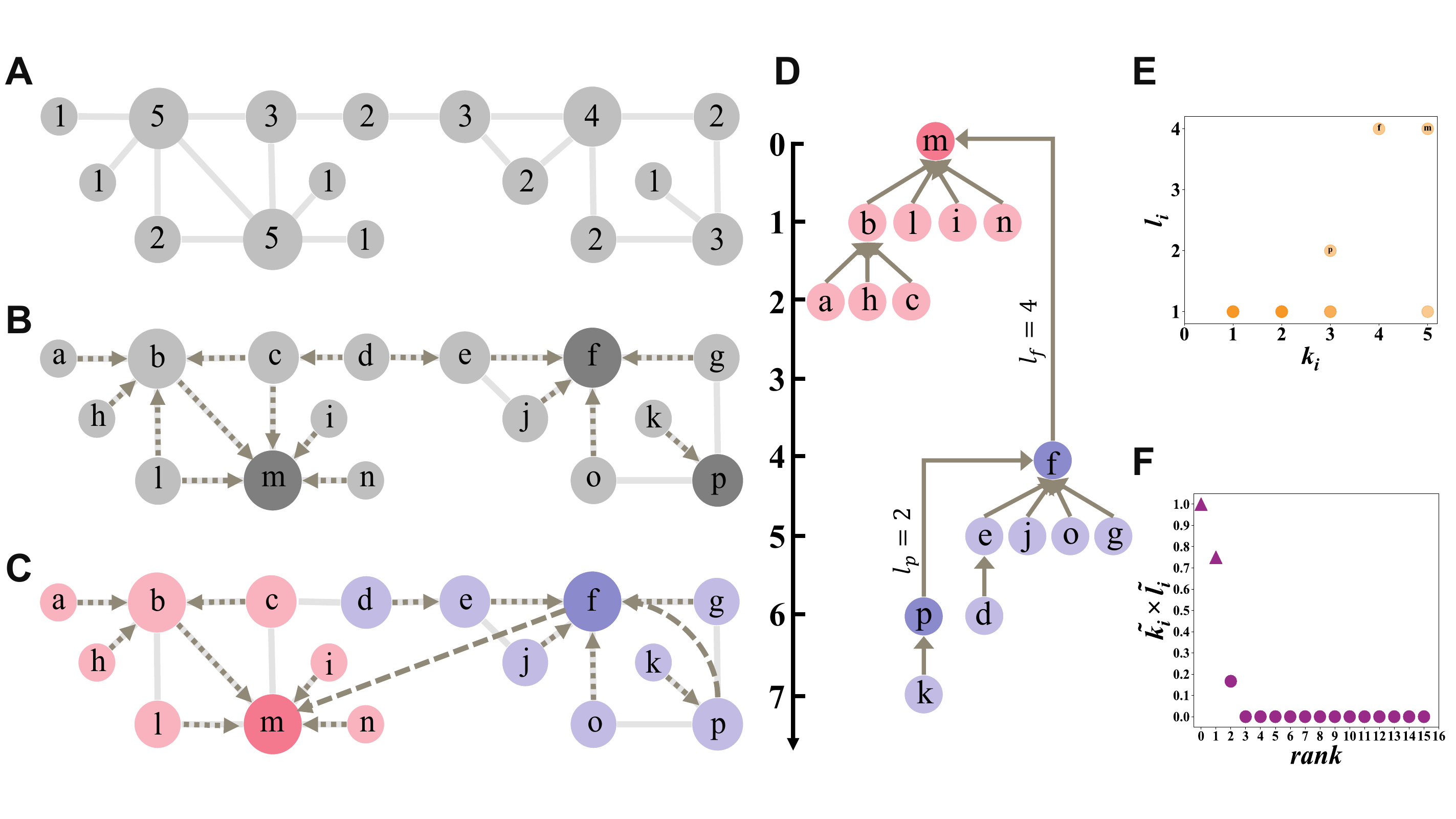}
\caption{\label{fig:example} 
\textbf{Schematic illustration of the Local Search (LS) algorithm.} 
(\textbf{A}) An example network where digits on nodes and size of nodes indicate the degree. 
(\textbf{B}) The identification of local leaders based on local dominance by creating a
forest of DAGs as indicated by short dashed directed edges. 
For each node $u$, it points to any adjacent neighbor $v$ with $k_v \geq k_u$ and $k_v=\max\{k_z | z \in \mathbf{V}(u) \}$, where $\mathbf{V}(u)$ is the set of neighboring nodes. In this example, nodes are traversed by their lexicographical order, when node \textit{b} is traversed, it points to \textit{m} as $k_m=\max\{k_z | z \in \mathbf{V}(b) \} \geq k_b$; later, when \textit{m} is traversed, it has no out-going link, and so \textit{m} is identified as a local leader: it does not point to any of its followers and its remaining neighbors all have smaller degrees. When there are more than one neighbor with the same largest degree, more than one directed edge is temporarily added, e.g.,\ node \textit{c} points to both \textit{b} and \textit{m} as $k_b=k_m=\max\{k_z | z \in \mathbf{V}(c) \} \geq k_c$; nodes \textit{d} and \textit{l} also have more than one outgoing link. 
The local leaders, which are potential community centers, are $f$, $m$, and $p$ (indicated by dark grey color). 
(\textbf{C}) Each node randomly retains just one out-going edge shown as a short dashed directed edge (e.g., \textit{c} can point to \textit{b} or \textit{m} with an equal probability, similarly for \textit{l} and \textit{d}). 
Then, for each local leader $u$, a local-BFS is performed to find its nearest local leader with $k_v\geq k_u$, and the shortest path length on network $d_{uv}, \forall v$ is designated by $l_u$. Here, $p\rightarrow f$ with $l_p=2$, and $f\rightarrow m$ with $l_f=4$. 
In (\textbf{C}), short-dash arrows and long-dash arrows correspond to pure followers (whose $l_u=1$) and local leaders (whose $l_u\geq 2$), respectively. 
Each node has at most one out-going link $(u\rightarrow v)$, which can go beyond direct connections. The local leader(s) with the maximal degree has no out-going link (here node \textit{m}). 
(\textbf{D}) The corresponding tree structure formed by local dominance. The scale on the left is a visual aid for calculating $l_i$ between connected nodes in the DAG. 
(\textbf{E}) The scatter plot of $k_i$ and $l_i$ for all nodes. Community centers are of both a larger degree $k_i$ and a longer $l_i$. 
(\textbf{F}) The decision graph for quantitatively determining community centers (indicated by triangles) based on the product of rescaled degree $\tilde k_i$ and rescaled distance $\tilde l_i$ (see more details in Supplementary Note 1.2). Community centers can be detected by a visual inspection for obvious gaps or sophisticated automatic detection methods. Here, two centers, nodes \textit{m} and \textit{f}, are identified. The color of nodes in (\textbf{C}) and (\textbf{D}) represents the community partition, and community centers are highlighted by a darker hue of the same color.   
}
\end{figure}

\section*{Results}
\subsection*{Local search algorithm}
Cluster analysis and community detection share many conceptual similarities, but often have a contrasting focus.
Cluster analysis puts emphasis on the center of a cluster \cite{kaufman2009finding,rodriguez2014clustering}, while community boundaries often play a more predominant role in community detection \cite{zitnik2018prioritizing}. Community centers can be inferred from some community detection algorithm outputs, for example, the nodes associated to the largest absolute weights of the leading eigenvector of the modularity matrix, or exhibiting a higher density of connections inside the communities, are deemed to be community centers, core members or provincial hubs \cite{newman2006modularity,guimera2005functional}. 
But centers are only a by-product of those algorithms, rather than at their core of methodologies. 

The approach that we propose here is explicitly focusing on community centers to identify clusters, which is motivated by the existence of underlying asymmetries between nodes \cite{serrano2009extracting,stanoev2011influence,lee2021uncovering}, the concept of \textit{local leaders} \cite{blondel2008local} in networks and borrows ideas from density and distance based clustering algorithms on vector data \cite{rodriguez2014clustering}. 
In our local searching (LS) algorithm, the local dominance refers to a leader-follower relation, and we pose a further restriction that each node has eventually at most one out-going link pointing to its leader. We hypothesise that communities are organized around centers that are nodes with both a dominant position at its neighborhood (e.g., has a larger degree, or other centrality measures, than its neighbors) and distant enough from other potential centers. Then based on community centers, partition is naturally ensuing. The process of our LS algorithm involves four steps: 

Firstly, we calculate the degree $k_u$ of each node (see digits in Fig. \ref{fig:example}A). 
Secondly, we point each node $u$ to its largest-degree-neighbor $v$ if this neighbor is no smaller than itself on degree (i.e., $k_v\geq k_u$ and $k_v=\max\changeII{\{}k_j|j\in V(u)\changeII{\}}$, where $V(u)$ is the set of neighboring nodes of $u$).
Nodes with in-going edge(s) and no out-going edge are \textit{local leaders} \cite{blondel2008local} that dominates its vicinity (see nodes \textit{f}, \textit{m}, and \textit{p} in Fig. \ref{fig:example}B). Such local leaders are like rich-among-poor and are potential community centers. 
Thirdly, for each local leader $u$, we use a local breath-first searching (LBFS) to find it a nearest local leader $v$ with $k_v\geq k_u$ and record its shortest path length to node $v$ as $l_u=d_{uv}$, which is larger than one (see long-dash arrows in Fig. \ref{fig:example}C, and see Fig. \ref{fig:example}D for a better extracted local dominance relation, which is in a reverse direction of arrows).  
The LBFS process stops after finding such a local leader $v$, which is the reason why we call it local BFS, so it generally searches a small region and does not traverse the whole network. For local leader(s) with the maximal degree, we do not perform such a LBFS, and directly assign the maximal $l_u$ of other local leaders (see mathematical descriptions in Methods). 

After performing LBFS to all local leaders except the maximal node, we can determine community centers according to degree $k_u$ and distance along the local dominance relation $l_u$ 
(in the network in Fig. \ref{fig:example}, nodes \textit{f} and \textit{m} stands out as centers, which have both a large $k_u$ and $l_u$, see Fig. \ref{fig:example}E). Note that for nodes except local leaders, their $l_u=1$. By multiplying normalized $k_u$ and normalized $l_u$, we can better quantitatively identify community centers via a notable gap between candidates (see Fig. \ref{fig:example}F, and details in Methods and Supplementary Note 1).  
Lastly, after the identification of community centers, the group label can be assigned to its followers along the local dominance relation (i.e., the reverse direction of arrows in Fig. \ref{fig:example}C) in one single step. 

We name our framework as local searching (LS) algorithm, since it only require local information of nodes and rely on efficient LBFS processes for local leaders, which takes up a very small fraction of the whole network (see Supplementary Table 1). The identification of local dominance relation is quite resilient to missing and noisy links. 
Our LS algorithm is of a linear time complexity \changeII{in terms of the number of edges} (see Methods and Supplementary Table 1) and is in no need of iteratively optimizing an objective function that relies on a global null model in other state-of-the-art methods \cite{blondel2008fast,rosvall2008maps,peel2017ground,duch2005community,mclachlan2007algorithm} or resorting to spreading dynamics \cite{raghavan2007near,frey2007clustering}. 
In addition, our LS algorithm is also capable of identifying multiscale communities structure, as local dominance relation also provides us hierarchies between communities via asymmetric relationship between communities centers.
The strength of our framework is demonstrated on several classical challenging synthesized test cases (see Figs. \ref{fig:synthesized}-\ref{fig:multiscale}) and empirical network datasets with ground-truth community labels (see Table \ref{tab:net-F1}). 
Finally, we also show how it provides a connection to clusters in a metric space, and our LS algorithm outperforms current state-of-the-art unsupervised both clustering and community detection methods when applied to discretised vector data clouds (see Figs. \ref{fig:phaseTransition}-\ref{fig:clusterVector} and Table \ref{table:vec-F1}).


As the implementation of our algorithm was done in Python, we use the NetworkX package implementation of the Louvain algorithm, our main point of comparison, as they are both of a linear time complexity, to obtain fair comparison for running time (see Table \ref{tab:net-F1}), and we also compare with a broader range of popular community detection algorithms on partition performance (see Table \ref{tab:comparisons}), some of which are slower but more accurate ones. Our LS algorithm still ranks first or second on the partition performance for five out of seven networks. 
More details of our LS algorithm can be found in Methods and Supplementary note 1.


\subsection*{Synthetic networks}

\begin{figure}[]\centering
\includegraphics[width=0.6\linewidth]{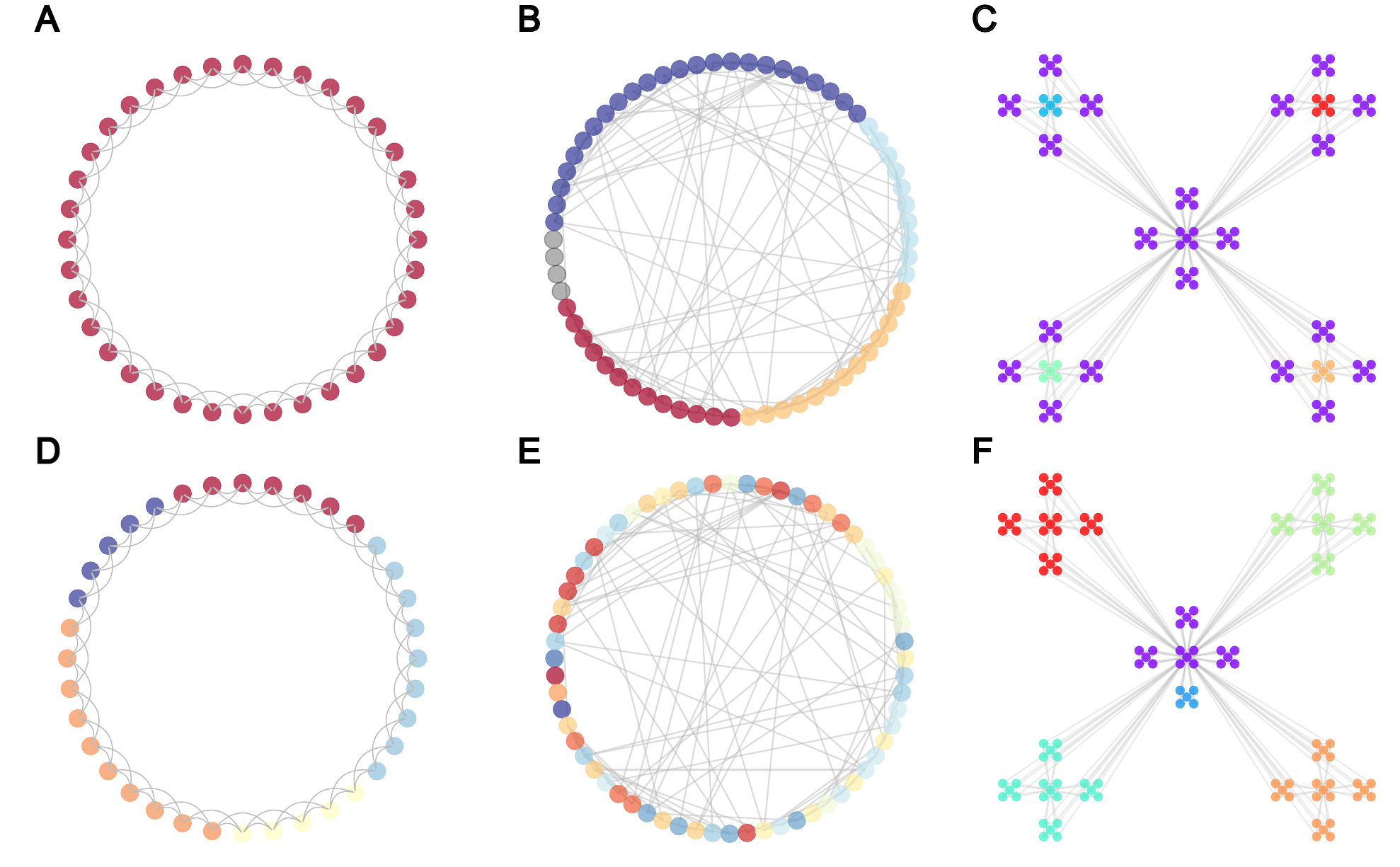} 
\caption{\label{fig:synthesized} 
	\textbf{Community partitions by the LS and Louvain algorithms on synthesized networks with different strength of heterogeneity}. 
	The heterogeneity increases from left to right. The color of nodes denotes the community membership. In a strict homogeneous regular network ($N=36, \langle k\rangle=4$), all nodes are identical, (\textbf{A}) only one community is detected by the LS algorithm (see Supplementary Fig. 1 for more details); (\textbf{D}) by contrast, the Louvain algorithm detects five communities by optimizing modularity. In an Erd\H{o}s-R\'enyi random network ($N=64, \langle k\rangle=4$), there may exist some communities due to randomness \cite{reichardt2006networks}, (\textbf{B}) the LS algorithm detects fewer communities compared to (\textbf{E}) the Louvain algorithm (see Supplementary Fig. 2). In a Ravasz-Barab\'asi network \cite{ravasz2003hierarchical} which displays stronger heterogeneity, (\textbf{C}) the LS algorithm groups all first-level nodes and all sixteen second-level peripheral clusters into one community, and four small communities emerge (see Supplementary Fig. 4 for more details); (\textbf{F}) the Louvain algorithm partitions each second-level branching as a separate community and misclassifies a first-level peripheral cluster into its own community, a result of traversal order and modularity optimization process in the Louvain algorithm.}
\end{figure}

Here, we use well-known benchmark networks to illustrate how the LS method functions and in which situations it performs well. 
For illustration, we mainly contrast the results obtained by the LS method to those obtained by the Louvain method \cite{blondel2008fast}, which is widely applied due to its good performance and high efficiency. In addition, both algorithms have a linear time complexity, and thus the comparison on performance between them are more fair. 
We first look at a circular regular network, where all nodes are equivalent and thus no community structure should be discovered. LS correctly identifies a single community (Fig. \ref{fig:synthesized}A), by contrast, modularity forces community structure to exist and finds five communities (Fig. \ref{fig:synthesized}D). 
Let us look in detail at the reason why LS finds a single community. First, each node will point to all its adjacent neighbors as they all have the same degree, and since node are sequentially traversed and they will not point to their followers, loops cannot be formed, see Supplementary Fig. 1C and Supplementary Note 1.1.1 for a proof. After all nodes have been considered, each node will only keep one outgoing link with an equal probability, and eventually a tree structure will be formed. Because of the homogeneity of the graph, the tree only allows the identification a single community center and therefore of a single community. Because all nodes are equivalent, the labeling and thus order in which they are visited, is irrelevant. We note that in the case of a clique, an extreme case of regular network, the mapping of local hierarchy can yield a range of structure from a chain to a star structure, see Supplementary Fig. 1B for more details. In all cases, only one center is identified. By contrast, the Louvain method would partition a homogeneous regular network into several communities by optimizing modularity (see Fig. \ref{fig:synthesized}D). 

Our second application focuses on Erd\H{o}s-R\'enyi (ER) random graphs, which is still relatively homogeneous though not strictly homogeneous. While in the limit of an infinite random graph no community structure exists, in finite-size ER graphs, fluctuations may create spurious community structures \cite{reichardt2006networks,fortunato2016community}, as well as weak or spurious hierarchies between nodes. In this example, the LS method detects fewer communities than the Louvain algorithm, see Fig. \ref{fig:synthesized}B and Fig. \ref{fig:synthesized}E. 
In ER random networks, the degree distribution is relatively restricted around its average, but the system nonetheless exhibits fluctuations in the degrees. Large degree nodes are more likely to connect to each other, as the connection probability between them, $k_ik_j/2E$, are among the highest ones. 
When two large nodes are connected, there will be a directed out-going link pointing from one node to the other, making one of them a follower. Thus the LS method detects fewer communities. 
On the other hand, when we fix the size of the network and increase the connection probability $p$, the number of communities detected by the Louvain algorithm also decreases but it consistently finds more communities than the LS method (see Supplementary Fig. 2). 
In addition, we are able to detect isolated nodes as noise (see grey nodes in Fig. \ref{fig:synthesized}B), as these nodes are of a small degree but infinite $l_i$. 

We also consider an extension of the ER random graph model, the stochastic block models shown in Supplementary Fig. 3 and discussed in Supplementary Note 1.1.2.
For random networks generated by stochastic block model \cite{holland1983stochastic,newman2016structure,schaub2020hierarchical} 
with two blocks, when the inter-connection probability is zero, $c_{out}=0$, the Louvain algorithm detects two communities that align with ground truth, but this is a reflection of the resolution limit \cite{fortunato2007resolution}, as when analyzing each community (each ER graph), it may partition it into more than ten communities (see Supplementary Fig. 2). By contrast, the LS algorithm still detects as many community centers as when looking at each individual random network. This result can be understood by the local nature of the algorithm, where the structure of one disconnected cluster does not affect the communities found in the other and thus LS algorithm does not suffer from resolution limit.  
And when fixing the intra-connection probability $c_{in}$ and increasing $c_{out}$, the boundary of the two communities becomes blurred. We find that when slightly increasing $c_{out}$, the number of community partitions given by the Louvain algorithm increases drastically (see Supplementary Fig. 3B).  
By contrast, the $F_1$-score of the LS algorithm is relatively stable, though not too high, and outperforms Louvain when $c_{out}$ is larger (see Supplementary Fig. 3).

Finally, we consider a hierarchical benchmark, the Ravasz-Barab\'asi network model \cite{ravasz2003hierarchical} with two layers, which naturally provides a model with a hierarchy between the center and peripheral nodes. The clustering proposed by LS method groups explicitly reflects the hierarchical nature of the model by grouping first-level nodes and all sixteen second-level peripheral clusters into one community centerd at the original seed node, as it dominates their neighborhood; and four small communities emerge due to the existence of four centers, which have a degree larger than their neighbours and a longer path length to the original seed node, i.e., $l_i>1$, see Supplementary Fig. 4C for the decision graph that identifies community centers. 
The Louvain algorithm offers an alternative partitioning that ignores the hierarchical nature of the model and finds five communities of roughly equal size, and misclassifies a peripheral cluster as a separate small community (see Fig. \ref{fig:synthesized}F). This example is interesting in that the clustering provided by Louvain here provides a reasonable, yet alternative, answer that ignores one aspect of the data. 
This reminds us that different clustering methods rely on different underlying mechanisms and, as often occurs when using unsupervised methods, the outputs are rarely strictly right or wrong. The outputs should be understood not only in terms of the data, but of the methods as well. 
Still, it is worth noting that the Louvain algorithm misclassifies a first-level peripheral cluster into another community (see the blue cluster in Fig. \ref{fig:synthesized}F), due to the traversal order used by the algorithm and modularity optimization process (see Supplementary Note 1.1.3 for more details). 
When we further modify the network generated by the Ravasz-Barab\'asi model by adding a third-level branching to one of the second-level central cluster, and add noise in the connectivity to other second-level central clusters, the LS method still detects meaningful hierarchical structure, see Supplementary Fig. 4B and Supplementary Fig. 4D.

\subsubsection*{Detection of multiscale community structure} 

\begin{figure}[!htbp]\centering
    \includegraphics[width=\linewidth]{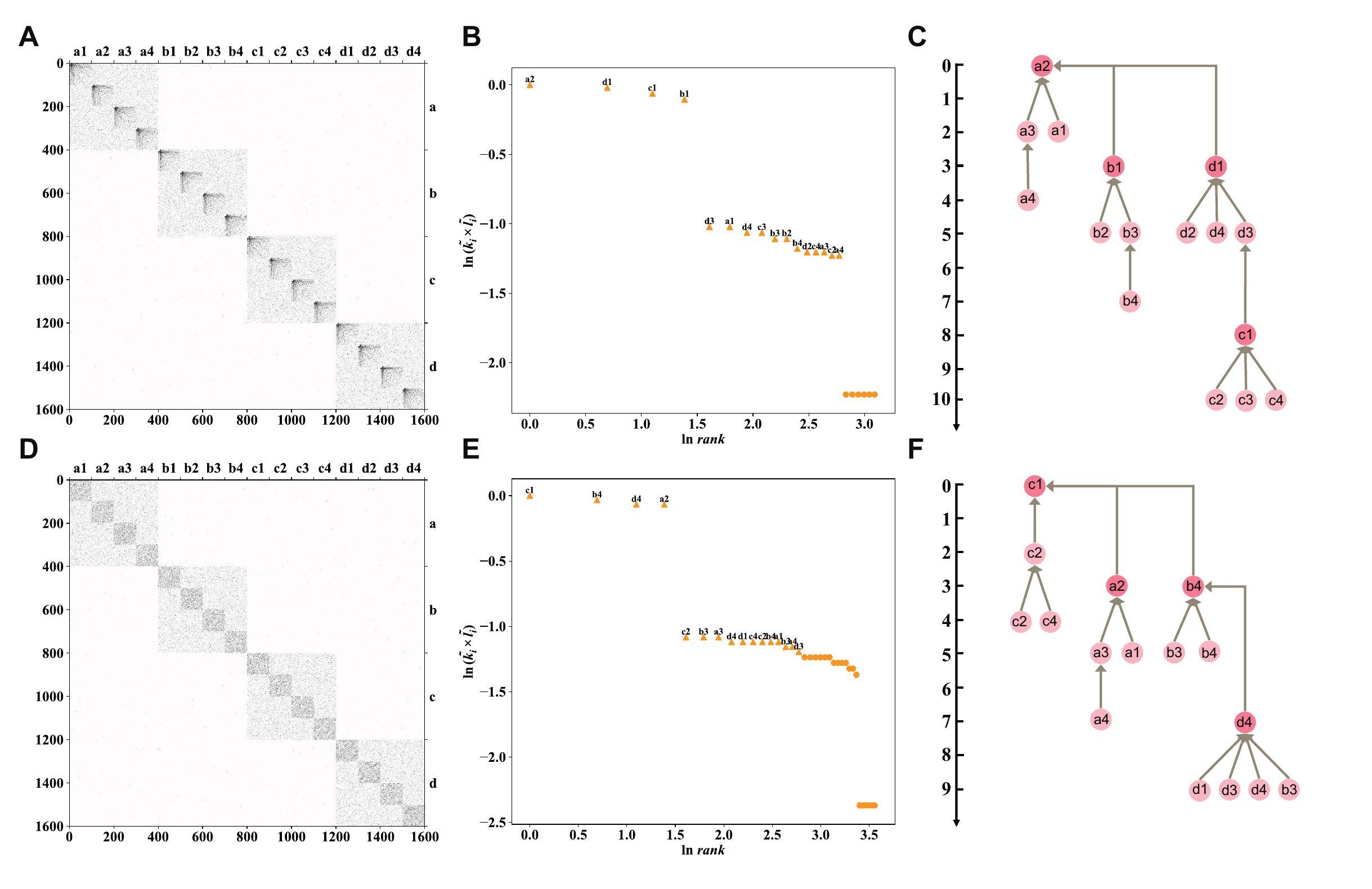} 
    \caption{\label{fig:multiscale} \textbf{Detection of multiscale community structure with different heterogeneity.} 
    The network in (\textbf{A}) comprise four top-level communities (labeled as a, b, c, and d) with 400 nodes each and an inter-connection probability $p_1=0.0002$, each of which further comprises four second-level communities with 100 nodes and $p_2=0.035$ (e.g., community c comprises c1, c2, c3, and c4). 
    The second-level communities are generated by the Barab\'asi-Albert model \cite{barabasi1999emergence} with $m=7$, which leads to an average degree $\langle k\rangle =14$. 
    (\textbf{B}) shows the decision graph for the LS method when analyzing the network in (A). 
    (\textbf{C}) displays the tree structure formed by the local dominance between identified centers of each community. For better clarity, community centers are named by the community label instead of the real index of the node, and we only show the tree structure of these centers. The height difference indicates the $l_i$ of the lower node. 
    (\textbf{D})-(\textbf{F}) is the same as (\textbf{A})-(\textbf{C}), with only changing the generation process of second-level communities to the Erd\H{o}s-R\'enyi random network with a connection probability $p=0.14$, which still leads to the same average degree $\langle k\rangle =14$. In such a setting, similar to SBM, nodes in the network are again relatively homogeneous. For better clarity, in (E) and (F) only top sixteen centers are labeled and their affiliation relation are visualized, and in total, LS detects 29 centers at the second-level for this network. 
    For the multiscale network in \textbf{A}, the LS method detects four top-level communities with $F_1=0.99$ and 16 second-level communities with $F_1=0.56$. For the network in \textbf{D}, the LS method detects four top-level communities with $F_1=0.89$ and 29 second-level communities with $F_1=0.29$. 
    In both cases, the Louvain algorithm only obtain four communities, which corresponds to the first-level ones, with $F_1$ equals $1$, however, it cannot detect second-level partitions. By comparing results in (A)-(C) and in (D)-(F), we can find that our LS algorithm works well on networks with stronger heterogeneity.  Results shown here correspond to just one realization, in multiple realizations, as every first- and second-level communities are equivalent, the label sequence in \textbf{B} and \textbf{E} and the tree structure in \textbf{C} and \textbf{F} may vary but have a consistent structure. 
}
\end{figure}

As partially reflected in the decision graph of the LS algorithm for the Ravasz-Barab\'asi network (see Supplementary Fig. 4C-D), 
the reliance on local dominance of our method to identify local leaders naturally lends itself to detect multiscale community structure \cite{sales2007extracting,clauset2008hierarchical,mucha2010community}. 
To illustrate this point, 
we generate a multiscale network made of two levels: four top-level communities with 400 nodes each and inter-connection probability $p_1=0.0002$, each top level community contains four second-level communities with 100 nodes each and $p_2=0.035$ \cite{sales2007extracting,clauset2008hierarchical}. Each second-level community is generated by the standard Barab\'asi-Albert model \cite{barabasi1999emergence} with $m=7$ that yields $\langle k \rangle=14$ (see Fig. \ref{fig:multiscale}A). 
The LS method correctly identifies two levels of community structure with a notable gap between first four top-level centers, which have similar $\tilde{k}_i\times\tilde{l}_i$, and other potential centers, as shown in Fig. \ref{fig:multiscale}B. 
Then taking the twelve subsequent centers, these sixteen centers together correspond to the sixteen second-level communities, 
and their affiliation within each top-level communities are correct (see the tree structure for local leaders in Fig. \ref{fig:multiscale}C). As all sixteen second-level communities are statistically equivalent, the directionality of community centers (Fig. \ref{fig:multiscale}C) is determined by fluctuations in the network generating mechanism. The partition obtained by the LS method has an $F_1$-score of 0.99 at the top level and of $F_1=0.56$ at the second level. Misclassifications at the second level mainly come from a relatively large inter-connection probability $p_2$, which blurs the boundary between communities. 
In comparison, the Louvain algorithm only detects four large communities that correspond to the top-level ones with $F_1$-score equals 1, but it cannot detect second-level smaller communities due to the resolution limit \cite{fortunato2007resolution}. 
This demonstrates the strength of the LS method on detecting smaller scale community structure.   

One reason that LS works on detecting multiscale structure resides in the fact that the average path length between nodes is governed by the connection probability \cite{evans2022linking}. The distance between nodes from different second-level communities within the same top-level community is on average shorter than the distance between nodes from different top-level communities, and thus the hierarchical structure is uncovered by the LS method. Another reason is the intrinsic heterogeneity in each second-level community. 

By contrast, when keeping the average degree and inter-connection probability ($p_1$ and $p_2$) the same, and replacing the second-level communities by ER random networks with $p=0.14$, which also yields $\langle k \rangle=14$ (see Fig. \ref{fig:multiscale}D), the whole network becomes more homogeneous (see Supplementary Fig. 5). 
In this case, the LS method can still detect four top-level communities (see Fig. \ref{fig:multiscale}E) but mis-identify some second-level communities (e.g., communities \textit{c2} and \textit{d1} are missing in this example, see Fig. \ref{fig:multiscale}F) and detect more smaller communities (29 second-level communities are detected instead of 16). 
The mis-identification of some second-level communities is due to the largest degree node $u$ in those ground-truth communities being directly connected to a node $v$ in other communities with $k_v\geq k_u$, and thus $u$ is considered as followers. 
This is more common in such a random setting, as there are more nodes with a relatively large degree beyond the reference value (i.e., the smallest degree of all of the largest node in each ground-truth second-level communities, see Supplementary Fig. 5 for more details). 
By contrast, in the scale-free case, there are fewer nodes beyond the reference value. 
For example, in the random multiscale network in Fig. \ref{fig:multiscale}D, the reference value is 34, and there are 60 nodes beyond it; in comparison, in the scale-free one, there are only 31 nodes beyond its reference value. 
The homogeneity makes the detection of such communities harder, if this minimum value become only slightly smaller, there will be much more nodes beyond the reference value in the random setting (see Supplementary Fig. 5B). 
Mis-affiliation, i.e., one local leader in community \textit{b3} follows the center of \textit{d4} instead of other centers in community \textit{d}, is also partially due to a similar reason and partially due to randomness. 
The discussion above also imply that the LS method would be vulnerable to targeted failure -- connecting two community centers would diminish one center as a follower and their corresponding communities merge as one (see Supplementary Fig. 24). 
In addition, due to randomness, two or more local leaders might emerge in the same second-level communities, which will lead to split of the community (e.g., there are two local leaders in communities \textit{c2}). 
These would constitute cases where the LS method is not appropriate.

\subsection*{Real-world benchmark networks}
\begin{table}[!htbp] \centering
\begin{tabular}{|c|c|c|c|c|c|c|c|}
\hline
         & $N$ & $E$ 
          &$\langle k \rangle $
          &$\langle d \rangle $    & $\langle CC \rangle $    & $\rho$ & $\alpha$ \\ \hline
Karate    & 34     & 78     & 4.59                       & 2.443                                              & 0.256 & -0.476        & 1.781              \\ \hline
Football\cite{evans2012football}  & 115    & 613    & 10.66                      & 2.397                                              & 0.407 & \ 0.162         & --                 \\ \hline
Polbooks  & 105    & 441    & 8.40                       & 2.841                                              & 0.348 & -0.128        & 1.791              \\ \hline
Polblogs  & 1,222  & 16,717 & 27.36                      & 2.747                                              & 0.226 & -0.221        & 1.415              \\ \hline
Cora      & 2,485  & 5,609  & 4.08                       & 5.738 & 0.117 & -0.055        & 1.645              \\ \hline
Citeseers & 2,110  & 3,668  & 3.48                       & 10.257                                             & 0.171 & -0.024        & 2.074              \\ \hline
PubMed    & 19,717 & 44,327 & 4.50                       & 2.764                                              & 0.060 & -0.044       & 2.227              \\ \hline
\end{tabular}
\caption{\changeII{\textbf{Basic statistics of networks.} $N$ is the number of nodes in the network, $E$ is the number of edges, $\langle k \rangle$ is the average degree of the network, $\langle d \rangle$ refers to the average shortest path length between all node pairs, $\langle CC \rangle$ refers to the average clustering coefficient, $\rho$ refers to assortativity, and $\alpha$ refers to the power-exponent of the degree distribution if it can be reasonably well fitted by a power law.}}
\label{statistics}
\end{table}

We now test the LS algorithm and demonstrate its strength on several empirical benchmark networks \changeII{(see Table \ref{statistics})} with known ground-truth community labels, see Table \ref{tab:net-F1}. 
\change{We chose to compare with the Louvain algorithm in Table \ref{tab:net-F1} because both algorithms are linear, making them well-suited for large-scale networks and facilitating a more meaningful comparison. 
And the Louvain method is the most widely used community detection algorithm implemented in most network packages. } 
LS is faster than Louvain for 7 of the 8 benchmarks. The speed advantage becomes more \changeII{noticeable} as the networks get larger (see Table \ref{tab:net-F1}). For example, for the DBLP network \cite{yang2013defining} with 317,080 nodes and 1,049,866 edges, our LS method takes 45 seconds, while Louvain takes 256 seconds.

The LS method is not only faster, but also classifies better than the Louvain algorithm measured by the $F_1$-score for 5 out of 7 examples with ground-truth community labels (see Supplementary Fig. 9 and Supplementary Note 2 for more details and discussions on the evaluation by $F_1$-score, \change{and we also make comparisons between \changeII{algorithms} on performance evaluated by conductance \cite{yang2013defining}, see Supplementary Table 4). 
In Table \ref{tab:comparisons}, we extend our comparison of the LS algorithm to include other popular algorithms 
with different perspectives on community detection, some of which are 
with greater accuracy albeit slower in implementation. 
For example, the GDG (geodesic density gradient) algorithm \cite{mahmood2016using} first embeds the network into vector space based on shortest path length between nodes and then applies an iterative clustering algorithm, which is similar to mean-shift algorithm \cite{comaniciu2002mean}, both of which are costly in computation, to obtain partitions of communities. GDG algorithm achieves the best performance in two out of seven networks, and has an obvious advantage over other methods on Citeseer (see Table \ref{tab:comparisons} and Supplementary Table 2 for more details). 
Another important type of method is inferential ones, which provides powerful tools without arbitrariness and further advances our understanding of community structure of networks \cite{peixoto2023descriptive}. Inferential algorithms, which generally \changeII{rely} on SBM as generative models, can explain the inability to detect communities in the very sparse limit \cite{decelle2011asymptotic}, help eliminate the resolution limit in Louvain algorithm and detect hierarchical structure of complex networks \cite{peixoto2014hierarchical}. Inferential methods can be adapted to different types of networks ranging from weighted networks \cite{aicher2015learning} to directed ones \cite{larremore2014efficiently} and hypergraphs \cite{contisciani2022inference}. However, inferential methods are generally computationally expensive \cite{peixoto2019bayesian,peixoto2014hierarchical}. The inferential algorithm described in Refs. \cite{peixoto2019bayesian,peixoto2014hierarchical} has a much better performance than the LS on \changeII{the} Football network (see Supplementary Table 3), whose degree distribution is quite homogeneous.  
Other algorithms typically only attain the first position in one out of seven networks. \changeII{While our LS algorithm} takes the lead in two out of seven networks and secures second place in an additional two. }
We note that the best performing algorithms are well distributed among the benchmarks, which reflects that real networks have generally different generating mechanisms that are better captured by some algorithm than others\cite{rosvall2008maps,peel2017ground}. 
\change{This suggests that achieving optimal performance across all scenarios is highly improbable \cite{peixoto2023implicit}, aligning with the No Free Lunch theorem \cite{peel2017ground}.} 
It is, however, interesting that LS consistently ranks first or second in the four out of seven benchmark networks, and is overall the best classifier, suggesting that the notions of local dominance, hierarchy and community centers are pervasive in real networks, whose degree distributions are generally heterogeneous. 
It is also instructive to understand why LS does not perform well on the Football network \cite{girvan2002community,evans2010clique,evans2012football}.
It is due to the fact that the Football network is fairly homogeneous, and we have already explained why the LS method does not perform well in this situation, see the subsection on multiscale community detection. There is also significant connectivity between the largest degree nodes in the ground-truth communities, thus some of them become direct followers to others and their communities are merged. If a portion of links between the largest degree nodes were removed, the partition given by the LS method would be much closer to the ground truth.

Targeted link removal and addition can significantly change the structure of a network and the outcome of community detection algorithms. The LS algorithm is not immune to that effect, as it relies on local leaders to separate communities, therefore, intentional targeted link addition between two community centers would make one of them a follower and lead to just a single community, which will dramatically reduce the performance of the classification. 
For example, if we connected the president and instructor in the Zachary Karate Club network, then the LS method only yields a single community (Supplementary Fig. 24), which is the case before the split \cite{zachary1977information}. This also lends us a way to identify critical links for merging or splitting communities \cite{palla2007quantifying}. 
The identification of local hierarchy is, on the other hand, more robust against links missing or adding at random, see Supplementary Figs. 22-23.

In addition, the number of communities detected by LS is also closer to the ground truth, see Table \ref{tab:net-F1}. 
For example, for the Zachary Karate Club network, Louvain detects four communities, while LS detects two, which is consistent with reality. As usually more potential centers can be detected in real networks, see Supplementary Fig. 6, and might correspond to meaningful multiscale structure. 
As for the Polblogs network, where LS finds three instead of two communities indicated by current labeling, and there is debate whether three groups should be considered as the ground truth (i.e., apart from liberal and conservative, there is a neutral community) \cite{adamic2005political}. This partially explains why the LS does not work that well on this example. This also reflects the importance and difficulty of obtaining ground-truth labels, if there are any \cite{peel2017ground}. 
Although the evaluation of the classification performance of an algorithm with a ground truth is standard practice \cite{hric2014truth}, establishing the ground truth for community assignment usually require detailed survey, which can be difficult for very large networks \cite{hric2014truth,newman2016structure}, and is usually regarded as distinct from metadata available \cite{newman2016structure,peel2017ground}. The choice(s) of the ground truth(s) \changeII{is} crucial and there might be ``alternative'' ground truth that emerge from unsupervised clustering analysis and are validated a posteriori. The notion of alignment between ground truth and structure is indeed crucial to obtain good clusters \cite{qian2021quantifying}. For example, in the well known Zachary Karate Club network \cite{zachary1977information}, the metadata of nodes can also be their gender, age, major, ethnicity, however, most of which are irrelevant to the community structure when interested in understanding the split of the club\cite{newman2016structure,peel2017ground}, but might be relevant to understand other type of community structure. 
\change{Apart from evaluations based on ground-truth labels, various evaluation criteria purely based on network structure (e.g., optimizing modularity, conductance, cut) have been proposed, however, they may deviate from the real generating process of networks and will not be suitable for all scenarios. For example, maximizing modularity cannot generate good partitions in ecological networks, as herbivores in the same community will not prey on each other, thus there are no dense connections within the same ecological community. In this sense, if there can be some ground-truth labels, using F1 score is a more objective evaluation. }

\begin{table}[!htbp] \centering
\begin{tabular}{|c|c|c|c|ccc|ccc|c|}
\hline
\multicolumn{1}{|l|}{\multirow{2}{*}{}} & \multirow{2}{*}{$N$} & \multirow{2}{*}{$E$} & \multirow{2}{*}{$N_c$} & \multicolumn{3}{c|}{Louvain}             & \multicolumn{3}{c|}{LS}      & \multirow{2}{*}{$\Delta t$ (ms)}            \\ \cline{5-10} 
\cline{8-10}
\multicolumn{1}{|l|}{}                  
&  &  &     
&\multicolumn{1}{c|}{$F_1$}  &\multicolumn{1}{c|} {$N_c$} & $t$ (ms)
&\multicolumn{1}{c|}{$F_1$}   &\multicolumn{1}{c|} {$N_c$} & $t$ (ms) &
\\ \hline

Karate  & 34  & 78  & 2                    
& \multicolumn{1}{c|}{0.63}  
& \multicolumn{1}{c|}{4}   & 8    
& \multicolumn{1}{c|}{\textbf{0.83}}   
& \multicolumn{1}{c|}{\textbf{2}}   & \textbf{6} & 2 \\ \hline

Football\cite{evans2012football} & 115 & 613 & 10       &\multicolumn{1}{c|}{\textbf{0.87}}  &\multicolumn{1}{c|}{\textbf{10}} & \textbf{18}            
&\multicolumn{1}{c|}{0.35} 
&\multicolumn{1}{c|}{6}  & 20 & -2 \\ \hline

Polbooks  & 105 & 441  & 3                    
& \multicolumn{1}{c|}{0.70}     
&\multicolumn{1}{c|}{5}  & 13 
&\multicolumn{1}{c|}{\textbf{0.80}} 
&\multicolumn{1}{c|}{\textbf{2}}  & \textbf{8} & 5\\ \hline

Polblogs  & 1,490   & 19,090   & 2          
& \multicolumn{1}{c|}{\textbf{0.85}} 
&\multicolumn{1}{c|}{9} & 328 
& \multicolumn{1}{c|}{0.69}      
&\multicolumn{1}{c|}{\textbf{3}}  &\textbf{212} & 116\\ \hline

Cora  & 2,708 & 5,429 & 7                    
&\multicolumn{1}{c|}{0.32}            
&\multicolumn{1}{c|}{28} & 380
& \multicolumn{1}{c|}{\textbf{0.33}}      
&\multicolumn{1}{c|}{\textbf{7}}  & \textbf{139} &241\\ \hline

Citeseers   & 3,264   & 9,072   & 6
& \multicolumn{1}{c|}{0.27}       
&\multicolumn{1}{c|}{35} & 384
& \multicolumn{1}{c|}{\textbf{0.45}}        &\multicolumn{1}{c|}{\textbf{7}}   & \textbf{131}  &253 \\ \hline

PubMed  & 19,717 & 44,327  & 3                  & \multicolumn{1}{c|}{0.20}     
&\multicolumn{1}{c|}{43} & 8,745
& \multicolumn{1}{c|}{\textbf{0.46}}         &\multicolumn{1}{c|}{\textbf{8}}  & \textbf{2,298} & 6,447\\ \hline

DBLP & 317,080 & 1,049,866 & --  & \multicolumn{1}{c|}{--}     
&\multicolumn{1}{c|}{220}  & 256,000
& \multicolumn{1}{c|}{--}         &\multicolumn{1}{c|}{8; 1859}  & \textbf{\textbf{45,000}} & 211,000 \\ \hline 

\end{tabular}

\caption{\textbf{Comparison between the LS and Louvain algorithms on networks with ground-truth community labels.} $N_c$ denotes the number of ground-truth communities in the network or identified by different methods, and $F_1$-score is a common performance measure in machine learning between predictions and ground-truth labels (see more details in Supplementary Note 2)
, and $t$ (ms) is the running time of the algorithm when implemented in Python. As there is no ground truth labels but only meta data for DBLP \cite{yang2013defining} (see Supplementary Note 2 for more discussions), we are unable to report $F_1$-score. As LS is able to detect multiscale structure, we report the number of communities detected with notable gaps: 8 large communities, 1859 smaller communities. 
Both the Louvain and LS algorithm are of linear complexity in time, and our LS method is faster. In addition, the LS method performs better in most cases. The algorithm with a better performance is highlighted in bold. Comparisons with a broader range of classical community detection algorithms are shown in Table \ref{tab:comparisons}.} 
\label{tab:net-F1}
\end{table}

\begin{table}[!ht]\centering
\resizebox{\textwidth}{!}{\begin{tabular}{|c|c|c|c|c|c|c|c|c|}
\hline
& Karate & Football\cite{evans2012football} &Polbooks  & Polblogs  & Cora & Citeseers & Pubmed & Time Complexity           \\ \hline
Spin glass \cite{reichardt2006statistical}          & 0.61     & \underline{0.92}            & 0.62          & 0.88 & \underline{0.33} & 0.22               &0.21  &  NA \\ \hline
GN \cite{girvan2002community} &  0.59  & 0.84             & \textbf{0.80} & 0.74                & 0.32                & 0.23   & 0.18     & O$(N^3)$        \\ \hline
\change{GDG} \cite{mahmood2016using}  & \textbf{0.91} &0.60 &0.75 &0.66 &0.29 &\textbf{0.63} &0.19 & O$(dtN^2)$ \\ \hline
Walktrap \cite{pons2005computing}           & 0.51    & 0.88             & \underline{0.79}          & 0.88                & 0.29                & 0.14       & 0.16    & O$(N^2\log N)$    \\ \hline 
Spectral \cite{newman2006modularity}  & 0.62   & 0.54             & 0.70          & \underline{0.89}                & 0.32                & 0.25              & \textbf{0.46}  & O$(N^2\log N)$\\ \hline
\change{Inferential} \cite{peixoto2019bayesian,peixoto2014hierarchical}  &0.65 &0.87 &0.77 &0.32 &0.31 &0.40 &0.07 &O$(N^2\log N)$ \\ \hline
Fastgreedy \cite{clauset2004finding}         & 0.75     & 0.56           & 0.78          & \underline{0.89}                & \textbf{0.39}       &0.28 & \underline{0.32} & O$(N\log N)$\\ \hline
Infomap \cite{rosvall2008maps}           & 0.76    & \textbf{0.96}            & 0.69          & 0.80                & 0.07                & 0.04        & 0.01     & O$(N\log N)$   \\ \hline
LPA \cite{raghavan2007near}                 &\underline{0.88}   & 0.79    & 0.69          & \textbf{0.91}       & 0.22                & 0.11           &0.18   & $\sim$ O$(E)$  \\ \hline
Louvain \cite{blondel2008fast}             & 0.63       & 0.87        & 0.70          & 0.85                & 0.32                & 0.27         & 0.20     & $\sim$ O$(E)$  \\ \hline
LS                 &0.83   & 0.35 & \textbf{0.80} & 0.69                & \underline{0.33}                & \underline{0.45}      & \textbf{0.46} & O$(E)$\\ \hline
\end{tabular}}
\caption{\change{\textbf{Comparisons with classical community detection algorithms on real networks with ground-truth community labels.} The algorithm with the highest $F_1$ score is highlighted in bold, and the second highest one is highlighted by underline. Overall, our LS algorithm have a pretty good performance, it is ranked first in two out of seven networks and ranked second in another two when compared to other popular algorithms, some of which are slower but more accurate ones. And our algorithm is the fastest one. For the GDG algorithm, $d$ is the dimension of embedding space, and $t$ is the number of iterations.} 
}
\label{tab:comparisons}
\end{table}

\subsection*{Applications to urban systems} 
\begin{figure}[!ht]\centering
\includegraphics[width=\linewidth]{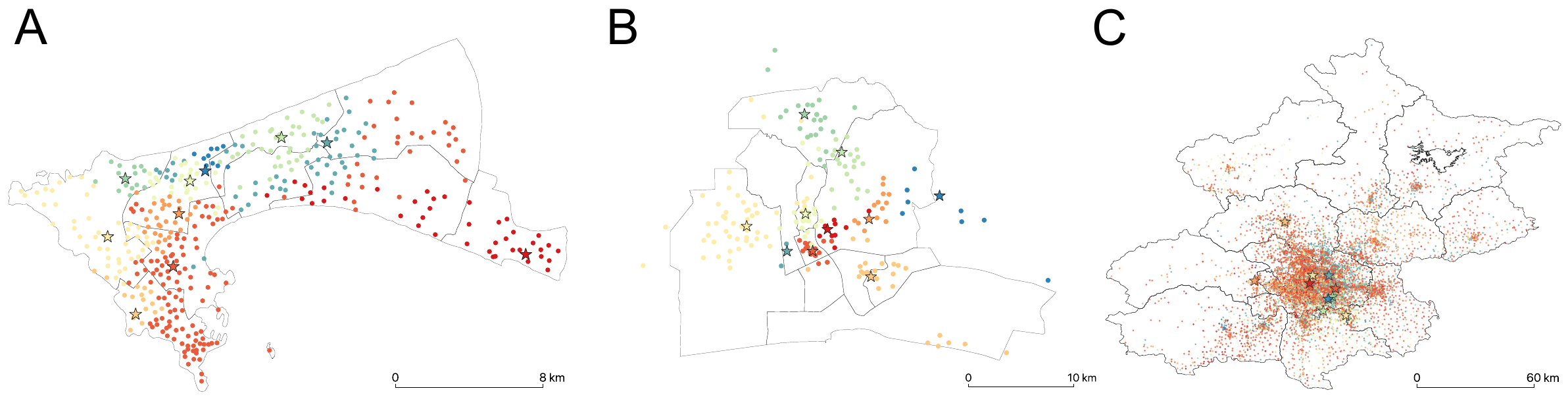}
\caption{\label{fig:city}\textbf{The community structure detected by our LS algorithm on mobility flow networks in three diversified cities across continents.} (\textbf{A}) Dakar in Senegal, Africa. (\textbf{B}) Abidjan in C\^ote d'Ivoire, Africa. (\textbf{C}) Beijing in China, Asia. Each dot represents a location, which corresponds to a region by Voronoi tessellation according to cellphone towers. Communities are indicated by different colors, and their centers are marked as stars. The decision graphs are shown in Supplementary Fig. 8.}
\end{figure}

Our final example of real-world networks is to uncover the structure of spatial interactions in cities. It also showcases the capacity of LS to adapt to weighted networks, with node degree replaced by the node strength and the least weighted shortest path, where the distance between two adjacent nodes is the reverse of the volume of mobility flow. 
Many cities have or will evolve from a monocentric to a polycentric structure \cite{louf2013modeling}, which can be inferred from the patterns induced in human mobility data. 
We use human mobility flow networks derived from massive cellphone data at the cellphone tower resolution with careful noise filtering and stay location detection \cite{alexander2015origin,ccolak2015analyzing,ccolak2016understanding} for three cities in different continents: Dakar \cite{dong2016population}, Abidjan \cite{li2017effects,liu2022revealing}, and Beijing\cite{xu2017clearer,xu2019unraveling} (see references here and Supplementary Material of ref. \cite{liu2022revealing} for more details on obtaining the mobility flow network from cellphone data). The LS algorithm can detect both communities with strong internal interactions and meaningful community centers, see Supplementary Fig. 8 for the decision graph. 
We find that for the smaller cities Dakar and Abidjan, communities are more spatially compact, while in the larger city, Beijing, they are more spatially mixed, see Fig. \ref{fig:city}. This indicates that in Beijing, interactions are less constrained by geometric distance, which might be due to a more advanced transportation infrastructure and a superlinearly stronger and diversified interactions tendency in larger cities \cite{schlapfer2014scaling,li2017simple,liu2022revealing}.  
In addition, the identified community centers correspond to important interaction spaces in cities, see Fig. \ref{fig:city}. For example, in Beijing, the top three centers are The China World Trade Center in Chaoyang District, the Zhongguancun Plaza Shopping Mall in Haidian District, and Beijing Economic and Technological Development Zone in Daxing District. In Abidjan, LS detects the Digital Zone, local mosques, and markets as centers. In Dakar, a university and some mosques are detected.


\subsection*{Clustering vector data via the LS algorithm} 
Community detection and vector data clustering share many similarities, but are often considered separately and having contrasting focus. 
Our use of local leaders identified by local dominance was directly inspired by the concept of the center of a cluster, which is characterised by a higher centrality measure in its vicinity/neighborhood (e.g., density or degree) and a relatively long distance (i.e., a large $l_i$) to the nearest object with a large centrality. 
Local dominance concretely and explicitly identifies fundamental asymmetric leader-follower relation between objects, which naturally give rises to centers. This creates a direct link between the two viewpoints of network science and data science. It is therefore natural to ask whether LS would perform well, or even better, than vector data clustering methods on a discretised version of a data cloud.

\begin{figure}[!hb]\centering
\includegraphics[width=\linewidth]{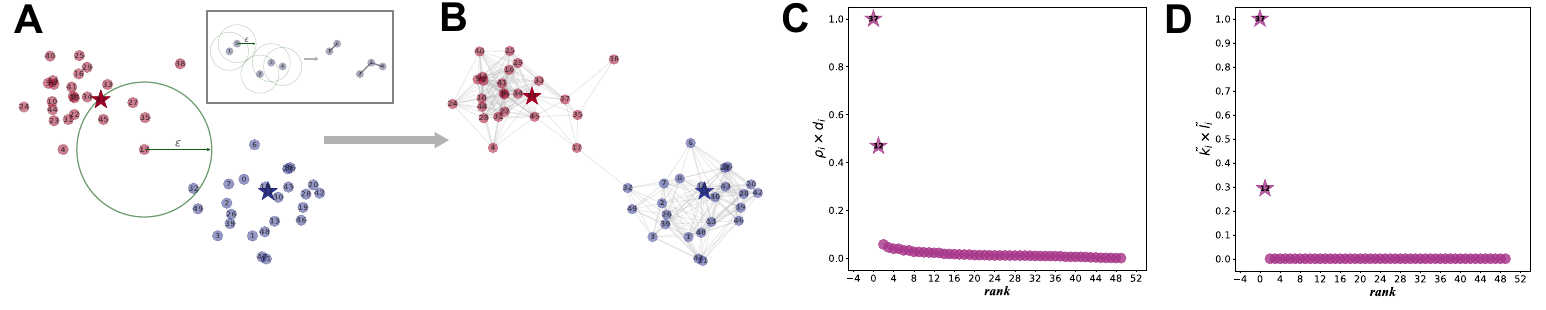}  
\caption{\label{fig:phaseTransition} 
\textbf{Conversion from vector data to a network via the $\varepsilon$-ball method and the analogy between the community centers of networks and the cluster centers of vector data.} 
(\textbf{A}) An example of data cloud and (\textbf{B}) its dicretised network representation by (\textbf{Inset}) the $\varepsilon$-ball method.
(\textbf{C}) The decision graph by the density and distance based (DDB) algorithm \cite{rodriguez2014clustering}. (\textbf{D}) The decision graph by the LS method. 
Cluster centers are data points of both a higher density $\rho_i$ than its neighbors and relatively far from other points with a larger density (i.e., a large $d_i$) \cite{rodriguez2014clustering}. 
The density $\rho_i$ of a data point $i$ is simply the number of nodes within a certain radius $\epsilon$, and it is equivalent to the degree of node $i$ in the corresponding network (i.e., $k_i=\rho_i$). 
The network constructing process is a coarse-graining and discretization process, where the absolute distance value is not preserved (e.g., in the Inset, $d_{32}>d_{34}$ for the original vector data, but $l_{32}=l_{34}=1$ in the network). The Euclidean distance between any data points is based on a global metric, but the topological path length between two nodes are based on a local metric. For example, $d_{24}$ is only slightly larger than $d_{34}$, but in the network, $l_{24}=2$ and $l_{23}=1$ (see the Inset); though $d_{21}\approx 2d_{23}$ according to global metric, node $2$ and node $1$ are not reachable in the network based on the local metric. 
Cluster centers identified by the DDB algorithm matches community centers identified by the LS method, which are all marked as stars. 
}  
\end{figure}

To cluster vector data with the LS method, we first need to discretise it into a network. 
Many methods exist to perform this task, including $\epsilon$-ball, k-nearest-neighbors (kNN) and its variants (such as mutual kNN, continuous kNN), relaxed maximum spanning tree \cite{qian2021geometric}, percolation or threshold related methods \cite{bullmore2011brain,hoffmann2020community}, and more sophisticated ones \cite{berenhaut2022social}. 
Here, we employ the commonly used $\epsilon$-ball method that sets a distance threshold $\epsilon$ and connects vectors, which become nodes, whose $\epsilon$-balls overlap, see Fig. \ref{fig:phaseTransition}A and inset. This process can be accelerated by using R-trees and are implemented in a time complexity of O($N\log N$)  \cite{guttman1984r,li2017simple} (see Supplementary Note 1.4). After traversing all nodes, a network encoding a geometric closeness within $\epsilon$ between nodes is obtained, see Fig. \ref{fig:phaseTransition}B. 
The $\epsilon$-ball method preserves spatially local information, e.g., the vector density in the metric space can be interpreted as degree in the constructed network, and coarse-grains continuous distance between objects into discrete values. This makes the determination of centers clearer (see Fig. \ref{fig:phaseTransition}C and D). 
The choice of $\epsilon$ influences greatly the structure of the network obtained, here we chose $\epsilon$ to be near the network percolation value to ensure a minimally connected graph \cite{bollobas2001random,bunde2012fractals,li2015percolation}, more details on determining $\varepsilon$ can be found in Supplementary Note 3.1.

\begin{figure}[!ht]\centering
\includegraphics[width=\linewidth]{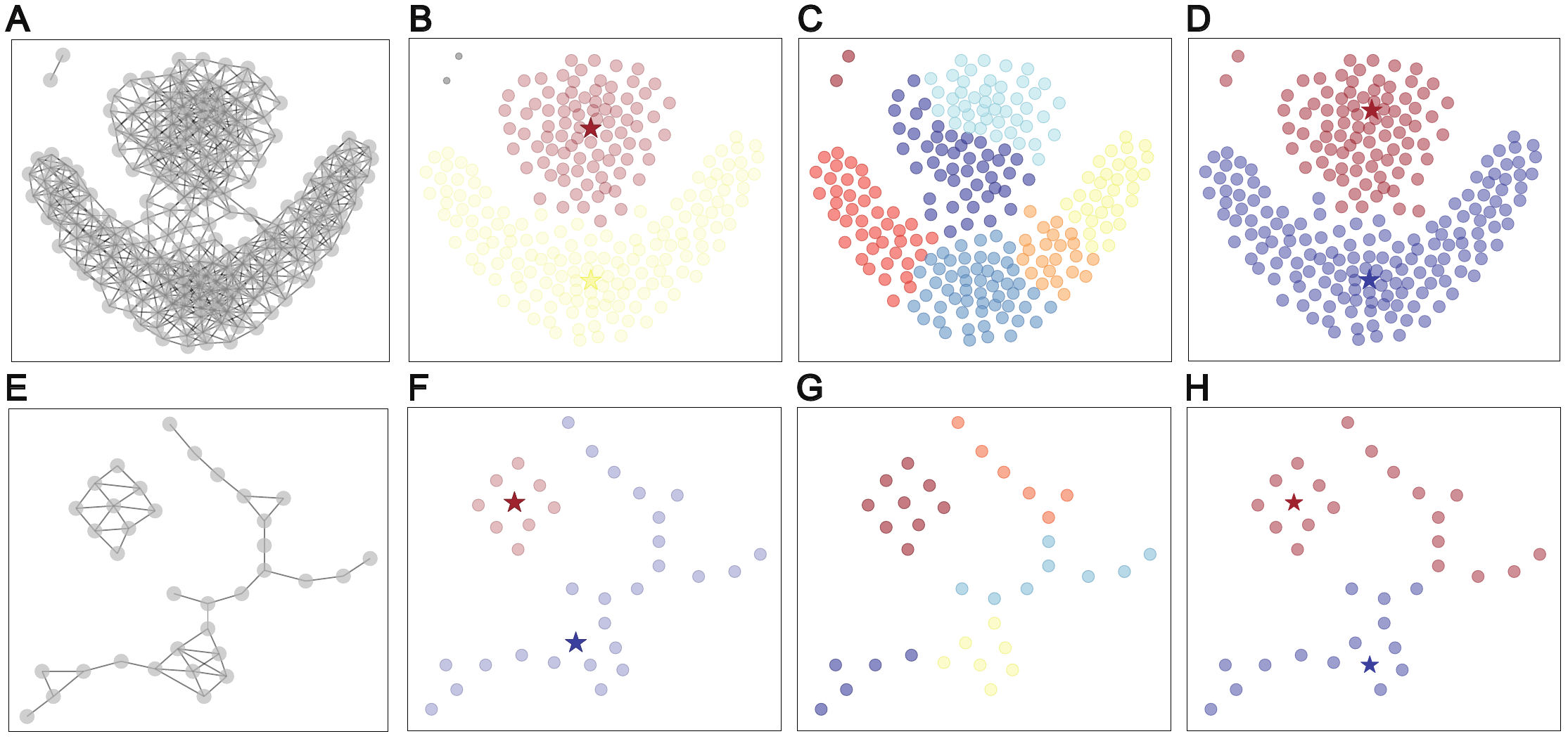}
\caption{\label{fig:clusterVector}
\textbf{Comparisons on the clustering performance between the LS, Louvain, and DDB algorithms for two dimensional benchmark vector data.} (\textbf{A}) and (\textbf{E}) represent networks constructed from vector data using the $\varepsilon$-ball method (see Supplementary Note 3.1 and Supplementary Fig. 10 for details on the network constructions). 
(\textbf{B}) and (\textbf{F}) show the result of the LS method  that correctly identify clusters that align with common consensus (see Supplementary Fig. 11 for more cases). In addition, LS can detect noisy points (marked in grey) that are of low degrees but long $l_i$. 
(\textbf{C}) and (\textbf{\changeII{G}}) show the partitions obtained from the Louvain method which are more fragmented than the LS results (see Supplementary Fig. 13 for more cases). 
(\textbf{D}\changeII{)} and \changeII{(}\textbf{H}) show the results obtained from the DDB method which provides correct partitions to most benchmark data, see (\textbf{D}) and Ref. \cite{rodriguez2014clustering} for other cases, but fails in the test case in (\textbf{E}), where both a low density manifold and a high density cluster exist, due to its local association rule \cite{pizzagalli2019trainable} being affected by a mixture of local and global metrics. 
LS and Louvain methods are performed on the constructed networks shown in \textbf{A} and \textbf{E}, and the DDB algorithm is performed on the original vector data. }
\end{figure}

Applying the LS algorithm on the constructed network for a series of well-known two dimensional benchmark data (Fig. \ref{fig:clusterVector}A and E, and Supplementary Fig. 10 for more cases), yields the expected clusters (Fig. \ref{fig:clusterVector}B and F, and Supplementary Fig. 11). By contrast, the Louvain algorithm 
generally obtains more and smaller clusters in a relatively fragmented way (Fig. \ref{fig:clusterVector}C and G, and Supplementary Fig. 13 for more examples) on the same networks. The reason is that the Louvain algorithm overlook the transitivity of local relations \cite{traag2019louvain}.
The state-of-the-art unsupervised clustering algorithm \textit{density and distance based} (DDB) \cite{rodriguez2014clustering} applied to the original vector data yields expected clusters in most cases, see Fig. \ref{fig:clusterVector}D and Ref. \cite{rodriguez2014clustering} for other examples. 
This confirms the universality of local hierarchy between objects and the analogy between our community centers and cluster centers. 
However, the DDB algorithm fails in the test case \cite{pizzagalli2019trainable} in Fig. \ref{fig:clusterVector}H due to a mixture of local and global metrics in this associate rule \cite{pizzagalli2019trainable}, which do not affect the LS method works (see Fig. \ref{fig:clusterVector}F).  
From a network perspective, certain dynamics can give rise to meaningful clusters with arbitrary shapes in metric space (e.g., synchronization or spreading dynamics are usually only possible along the manifold via local interactions but not through global ones). For example, different clusters in Fig. \ref{fig:clusterVector}E or Supplementary Fig. 11C,G,H might correspond to groups of fireflies that are only able to synchronize within the group rather than between groups, as their interaction range is usually limited. 
In the situations above, the distance measured by the local metric is more appropriate 
than the one measured by the global metric, see a more in depth discussion in Supplementary Note 3.2. 
The good performance of the LS algorithm on vector data resides in the correct identification of the local dominance, i.e., finding the centers, from the local metric.


In addition, we show that the LS method is robust against noisy data in different scenarios, see Supplementary Note 4.1 and Supplementary Figs. 23-24. 
Though less common when considering vector data, targeted addition of edges in a network that connect two cluster centers, explicitly brings two cluster centers closer to each other in the metric space and will distort the space,  whereas, conversely, the removal of links increases the distances between two objects.

\subsubsection*{The advantage of building networks for high dimensional vector data}

\change{We now show the advantage of combining the $\epsilon$-ball discretisation and community detection methods on clustering high-dimensional data sets.
Here, we use well-known benchmark datasets with very high dimensions}: the MNIST of hand written digits \cite{chatfield2011devil}, and Olivetti of human faces \cite{samaria1994parameterisation}, and show that our simple framework outperforms the-state-of-the-art DDB clustering algorithm \cite{rodriguez2014clustering}, see Table \ref{table:vec-F1}. Let us consider, for example, the Olivetti human face dataset, a challenging high dimensional dataset with small sample size. Each cluster obtained by the LS algorithm only contain images from a single individual, see Supplementary Figs. 17-19, simply based on Euclidean distance between images and without resorting to using complex image similarity measure. Moreover, it obtains a higher $F_1$-score than the DDB method. We note that for MNIST and Olivetti datasets, the Louvain algorithm has a higher $F_1$-score than LS, but identifies an inappropriately large number of clusters. The better performance of the Louvain algorithm lies in some subtle differences from clustering results obtained by the LS method (see comparisons between Supplementary Fig. 19A and Supplementary Fig. 19B for the Olivetti dataset with 100 images. The Louvain algorithm detects all images of the eighth person as one cluster, but the LS method classifies four images of the eighth person as another cluster). 

We conjecture that the conversion from vector data to a network is not merely a translation of the data, but a fundamental information filtering process that accentuates the prominence of local leaders and thus increases the strength of local hierarchy, which in practice turns out to be of great advantage to our framework for handling vector data with high dimensions. 
Constructing the network via $\epsilon$-balls is similar to a coarse-graining process: as long as two objects are close enough, the small differences in distances within $\epsilon$ are neglected. 
In addition, such a process also corresponds to subtracting irrelevant global information and puts the focus on similarity based on a local metric. 
Though there will be some information loss during the conversion from vector data to topological data, 
purely local information is enough to identify local dominance in the data. Not all information embedded in the vector data needs be utilized \cite{qian2021geometric}, sometimes too much information might complicate the process.
Although admitting asymmetric relations between objects would violate certain formal metric properties (distances are symmetric), it turns out to be an advantage for cluster analysis (see more discussions in Supplementary Note 3). 

\begin{table}[!hb] \centering
\begin{tabular}{|l|c|c|c|cc|cc|cc|}
\hline
\multirow{2}{*}{} & \multirow{2}{*}{$D$} & \multirow{2}{*}{$N$} & \multirow{2}{*}{$N_c$} & \multicolumn{2}{c|}{LS}                          & \multicolumn{2}{c|}{DDB}                        & \multicolumn{2}{c|}{Louvain}               \\ \cline{5-10} 
                  &                    &                    &                        & \multicolumn{1}{c|}{$F_1$}         & $N_c$       & \multicolumn{1}{c|}{$F_1$}         & $N_c$      & \multicolumn{1}{c|}{$F_1$}         & $N_c$ \\ \hline
Iris              & 4                  & 50                 & 3                      & \multicolumn{1}{c|}{0.73}          & 2  & \multicolumn{1}{c|}{\textbf{0.82}} & \textbf{3} & \multicolumn{1}{c|}{0.70}          & 8     \\ \hline
Wine              & 13                 & 178                & 3                      & \multicolumn{1}{c|}{\textbf{0.57}}          & \textbf{3}           & \multicolumn{1}{c|}{\textbf{0.57}}          & \textbf{3} & \multicolumn{1}{c|}{0.41} & 7     \\ \hline
MNIST             & 784                & 1,000              & 10                     & \multicolumn{1}{c|}{0.32} & \textbf{21}  & \multicolumn{1}{c|}{0.26}          & (10)   & \multicolumn{1}{c|}{\textbf{0.45}}          & 247   \\ \hline
Olivetti          & 10,304             & 100                & 10                     & \multicolumn{1}{c|}{0.74} & \textbf{14}  & \multicolumn{1}{c|}{0.64}          & (10)   & \multicolumn{1}{c|}{\textbf{0.78}}          & 32    \\ \hline
Olivetti          & 10,304             & 400                & 40                     & \multicolumn{1}{c|}{0.59} & \textbf{64} & \multicolumn{1}{c|}{0.49}          & (40)   & \multicolumn{1}{c|}{\textbf{0.68}}          & 112   \\ \hline
\end{tabular}
\caption{\textbf{Comparisons on the clustering performance between the LS, Louvain, and DDB algorithms for high dimensional vector data.} $D$ denotes the dimension of the dataset, $N$ denotes the number of objects, and $N_c$ denotes the number of clusters from the ground-truth or identified from algorithms. 
The hand-written figures in MNIST is of dimension 28$\times$28=784 pixels; and in Olivetti, the face image is of dimension 92$\times$112=10,304 pixels. The Olivetti dataset with $N$=100 comprises the first 100 images of 10 people from the original data set. The original dataset comprises 400 images of 40 different people. Our LS algorithm outperforms DDB in all high-dimensional and large-scale data sets except in Iris, whose dimension is quite low. Note that as the DDB algorithm does not have a clear recognition of the number of clusters (i.e., no clear gaps between centers in the decision graph) \cite{rodriguez2014clustering} for MNIST and Olivetti, the number of clusters identified by DDB are putative based on the ground truth (i.e., selecting the top ten or forty nodes in the decision graph), which is marked in brackets. Digits highlight in bold is the ones closest to the ground truth among all three algorithms.  
\label{table:vec-F1}}
\end{table}

\section*{Discussion}
Community detection and cluster analysis are analogous as both aim to group objects into categories based on some notion of similarity. 
In this work, we develop a fast and scalable community detection method based on the notion of a community center which echoes the commonly used concept of a cluster center.
The identification of community and cluster structures requires a heterogeneous system: uniformly distributed data points and strictly regular networks do not possess meaningful mesoscopic cluster structure. Heterogeneity leads to the emergence of more important loci in a data space, or central nodes in a network. The notion of center is pervasive in cluster analysis, but underused in community detection. We define community centers as local leaders that are both of a high degree, corresponding to a high density in cluster analysis, and relatively distant from other local leaders, corresponding to cluster separability. The nodes belonging to each community defined by their center are identified by basins of attraction \cite{sales2007extracting} based on the dominance existing between nodes, which indicates the asymmetric leader-follower relationship and defines a local hierarchy. 
While dominance is an explicit characteristic of edges in a directed network, it can be seen as an intrinsic hidden higher-order directionality between nodes even in undirected networks. 
The resulting local hierarchy reflects asymmetric interactions between objects inferred from the local connectivity of nodes that then naturally defines leaders and community affiliations, as well as hierarchies among communities. In addition, the position of local leaders and distribution of shortest path length $l_u$ between local leaders can be developed into some indicators for depicting network structure. And with the concept of local leaders and corresponding local hierarchy, automated discovery \cite{fortunato202220} and evolution dynamics of communities \cite{palla2007quantifying} can be ensuing studies. 

The local hierarchy structure is quite robust against random noise, and is based on local information. Moreover, in contrast to most state-of-the-art clustering and community detection methods, the LS method \changeII{does} not depend on the structure of the entire network as of most existing methods \cite{fortunato202220}. We are able to detect communities in a small region and avoid the computational burden of analysing the whole network \cite{fortunato202220}.
In cluster analysis, approximating similarity relations between objects by a distance matrix actually assumes that every object \changeII{is} in a direct relation with all others, which is also the case for modularity optimization \changeII{algorithms} that utilize a random null model, which also assumes that each node has a probability to interact with every other node \cite{fortunato2010community}. 
In addition, community detection methods also generally assume a mutual relation between objects, which is an important formal metric property and an implicit feature of an undirected connectivity matrix. Local hierarchy implicitly violates such an assumption, but it turns out that abandoning such a restriction gives better flexibility to the clustering method (see Supplementary Note 3 for more details). 
Finally, our LS algorithm is fast and scalable with a linear time complexity, which is crucial for analyzing large scale networks, and also performs well on most benchmarks, except the ones that do not possess the type of heterogeneity (e.g., football network \cite{evans2012football}) exploited by the LS method. 

Overall, the performance of the LS method is particularly good given its simplicity. On benchmark network models, it outperforms the currently most widely used community detection method, the Louvain modularity optimisation algorithm. The LS method consistently ranks higher than any other methods when the performance is averaged over several data sets, see Table \ref{tab:comparisons}. We have also shown that the LS method is naturally able to detect multiscale structure of communities in complex networks. This implies that while not necessarily identifying the partition defined by some existing ground truth, it finds a good approximation of it and the output can then be used as starting point for other slower but more accurate and dedicated community detection methods, offering a significant speed up. 

Given the similarity in spirit between LS and clustering methods, we applied LS to $\epsilon$-ball discretised version of benchmark vector data, both low and high dimensional. For low-dimensional data, we find it provides the expected clusters and outperforms Louvain modularity optimisation algorithm ran on the discretised data, which generally yields too many communities and performs poorly. LS also outperforms DDB, a state-of-the-art unsupervised clustering method, on some challenging cases in the presence of low-density manifolds. For high-dimensional data, LS still outperforms DDB, but not Louvain, although on closer inspection, Louvain obtains a better $F_1$-score, but suffers again from providing too many communities, outbalancing the advantage in $F_1$-score.

We hypothesise that the discretisation step of creating a network from vector data acts as a topological filter, which enhances the key property of the data that makes cluster detection work: the existence of well defined cluster centers and a clearer identification of local hierarchy. 
The performance of any community detection algorithm is going to be influenced by the discretisation method used, and more work is needed to understand the relationship between topological denoising and the performance of the community detection algorithms, as different community detection methods might respond differently to different discretisation schemes.

Another area for future work is to adapt LS to find ``halo'' nodes residing at the boundary of two or more communities (e.g., node \textit{d} in Fig. \ref{fig:example}), detect overlapping communities \cite{palla2005uncovering} potentially by producing line graphs \cite{evans2009linegraph,evans2009linegraphweighted,ahn2010link} or clique graphs \cite{evans2010clique}, and identify critical link responsible for the merging or splitting dynamics of communities \cite{palla2007quantifying}. 
Another point that could be improved is when two or more local leaders are equivalent on both degree and distance to a node. We currently assign it to a local leaders at random but we could look at other options. 

Finally, another possible direction for future research concerns the definition of dominance itself. In this article, it was built on a specific network property, the degrees of the nodes. For a weighted network, it would be appropriate to use strength rather than degree and we would retain all the benefits of the LS method. \changeII{Extending LS algorithm to directed networks is worth closer investigations in the future. In directed networks, two types of local leaders, the ``integrators’’ (determined by in-degree) and the ``influencers’’ (by out-degree), might be needed, which can lead to two types of clustering. The influence of edge directionality should be closely examined, as influence may propagate in the reverse direction of the directed edge. For example, on Twitter, information often flows from a user to their followers. Additionally, directionality affects the calculation of path lengths between nodes. } Apart from using degree, dominance could also be based on other node centrality measures but most of these require global network calculations which would slow the algorithm considerably. If Dominance was based on non-structural properties, such as numerical attributes for nodes already defined in the data, then the LS approach would still work well.

\section*{Methods} 
\subsection*{The Local Search (LS) algorithm}

Cluster analysis and community detection share many conceptual similarities, but often have a contrasting focus.
Cluster analysis puts emphasis on the center of a cluster \cite{kaufman2009finding,rodriguez2014clustering}, while community boundaries often play a more predominant role in community detection \cite{zitnik2018prioritizing}. Community centers can be inferred from some community detection algorithm outputs, for example, the nodes associated to the largest absolute weights of the leading eigenvector of the modularity matrix, or exhibiting a higher density of connections inside the communities, are deemed to be community centers, core members or provincial hubs \cite{newman2006modularity,guimera2005functional}. 
But centers are only a by-product of the algorithm, rather than at their core of methodologies. 

The approach that we propose here is explicitly focusing on community centers to identify clusters, which is motivated by the existence of underlying asymmetries between nodes \cite{serrano2009extracting,stanoev2011influence,lee2021uncovering}, the concept of local leaders \cite{blondel2008local} in networks and borrows ideas from density and distance based clustering algorithms on vector data \cite{rodriguez2014clustering}. We hypothesise that a community center is a local leader that is comparatively of a larger degree than its neighbors, thus ``dominating'' them, and is of a relatively long shortest-path distance to other local leaders. 

Our algorithm consists of four steps that we now detail. We start with an undirected network with $N$ nodes and $E$ edges, for example see Fig.\ \ref{fig:example}A. For better clarity, nodes are also labeled and traversed in lexicographical order (see Fig.\ \ref{fig:example}B).  

\begin{enumerate}
\item[Step 1] First, we calculate the degree $k_u$ of each node $u$ (see digits in Fig. \ref{fig:example}A), which is an operation of linear time complexity O$(E)$. 
\changeII{Our algorithm neglects self-loops in default, but if self-loops are meaningful for calculating degree of nodes, setting the input parameter ``self\_loop'' of the algorithm as True will increase the degree of nodes accordingly, and nodes with self-loops will not be considered as neighbors of themselves.}

\item[Step 2] Second, we traverse each node $u$ and point $u$ to any adjacent node $v$ with $k_v\geq k_u$ and $k_v= \max\{k_z|z\in\mathbf{V}(u)\}$ (i.e., $v$ has the largest degree in the neighborhood of $u$). For example, in Fig.\ \ref{fig:example}B node \textit{g} will point to \textit{f} instead of \textit{p} as $k_f>k_p>k_g$; and \textit{c} points to both \textit{b} and \textit{m} as $k_b=k_m=\max\{k_z | z \in \mathbf{V}(c) \} > k_c$. 
Note that a node cannot point to its follower, and since nodes are traversed in lexicographical order, when node \textit{b} is traversed, it will point to \textit{m} as $k_m=\max\{k_z|z\in\mathbf{V}(b)\}\geq k_b$. When \textit{m} is traversed, it will not point to any of its followers (e.g., \textit{b}).  
This process naturally avoids the creation of loops and ensure we only obtain directed acyclic graphs (DAGs), see Supplementary Fig. 1 and proof in Supplementary Note 1.1.1 for more details.
If such a $v$ does not exist, $u$ will not have any outgoing edge and will be identified as a local leader (see dark grey nodes \textit{f}, \textit{p}, and \textit{m} in Fig. \ref{fig:example}B). We denote the set of local leaders as $\mathbf{C}$. 

After traversing all nodes, for nodes with multiple out-going links, we randomly retain one (see only short dash arrows in Fig. \ref{fig:example}C for a possible mapping). Mathematically, we have obtained a forest of trees, where the root of each tree is a local leader, and is also a potential community center. For most nodes, except local leaders, this process identifies a local hierarchy (indicated by dash arrows), with an asymmetric leader-follower relation (see short-dash arrows in Fig. \ref{fig:example}B). 
This step is completed in O$(E)$.

\item[Step 3] Third, to identify the upper level for local leaders along the hierarchy, 
we use a local breadth-first search (BFS) starting from each local leader $u$ and stop the search when encountering the first local leader $v$ with $k_v\geq k_u$ and assign the shortest path length on the original network $d_{uv}$ to $l_u$, which is the length of the out-going link of node \textit{u}. Note that $l_u\geq 2$ for all local leaders, and all pure followers have $l_u=1$. 
For example, node \textit{p} is a local leader, in the second iteration of the BFS, it encounters another local leader \textit{f} with $k_f>k_p$. We stop the local-BFS and point \textit{p} to \textit{f}, and $l_p=d_{pf}=2$. 
Similarly, $f\rightarrow m$ and $d_{mf}=4$. 
The out-going link of local leaders goes beyond the direct connections in the original network (see long-dash arrows in Fig. \ref{fig:example}C).   

When there are several local leaders that have a no smaller degree than the local leader $u$ in the $l_u^{th}$ iteration, the largest one is chosen; 
if multiple nodes have the same largest degree, one is picked at random uniformly. 
For local leader(s) with the maximal degree in the whole network, denoted as $\mathbf{M}$, a subset of $\mathbf{C}$, there is no need to perform the BFS, and we directly assign $l_{x\in\mathbf{M}}=\max_{u\in\mathbf{C}\setminus \mathbf{M}}({l_u})$. 
 
Community centers can be easily identified as local leaders with both a large $k_u$ and a long $l_u$ (see Fig. \ref{fig:example}E), and naturally emerges from the rooted tree revealed by local dominance (see all dash arrows in Fig. \ref{fig:example}C and the explicit tree structure in Fig. \ref{fig:example}D). 
We use the product of rescaled degree $\tilde k_i$ and rescaled distance $\tilde l_i$ to quantitatively measure the ``centerness'' of each node (see more details and discussions in Supplementary Note 1.2). Community centers can be determined via visual inspection for obvious gaps or by, possibly, sophisticated automated detection methods for gaps in the future (see Fig. \ref{fig:example}F).  
For example, the community centers identified by the LS method in Fig. \ref{fig:example} are nodes \textit{f} and \textit{m}. 
In the Zachary Karate Club network, the identified community centers correspond to the president and the instructor, which is consistent with reality \cite{zachary1977information} (see Supplementary Fig. 6). 

Theoretically, this third step takes O$\left((|\mathbf{C}|-|\mathbf{M}|)\langle k\rangle^{\langle l\rangle_{\mathbf{C\setminus M}}}\right)$=O$\left((|\mathbf{C}|-|\mathbf{M}|)E\right)$, where $\langle k\rangle$ is the average degree of the network, $\langle l\rangle_{\mathbf{C\setminus M}}=\sum_{u\in \mathbf{C\setminus M}} l_u/(|\mathbf{C}|-|\mathbf{M}|)$, and the size of the set of potential centers $|\mathbf{C}|$ is usually much smaller than $N$ (see Supplementary Table 1). In practice, $\langle k\rangle^{\langle l\rangle_{\mathbf{C\setminus M}}}$ is bounded to be smaller than $E$ as it mimics a local-BFS process. As indicated by numerical results, even $(|\mathbf{C}|-|\mathbf{M}|)\langle k\rangle^{\langle l\rangle_{\mathbf{C\setminus M}}}$ is usually smaller than $E$ (see Supplementary Table 1). In addition, the local-BFS process can be simultaneously implemented for all local leaders in parallel to further speed up the algorithm in practice.

\item[Step 4] Finally, for all identified community centers, we remove their out-going links, if any. Community labels are then assigned along the reverse direction of directionality $u\leftarrow v$ from community centers. 
This step takes again a linear time O($N$). 

\end{enumerate}

Taken together, the time complexity of our LS algorithm is linear \changeII{in the number of edges}:  O$\left(E+(|\mathbf{C}|-|\mathbf{M}|)\langle k\rangle^{\langle l\rangle_{\mathbf{C\setminus M}}}+ N\right)=\Theta(E)$, which is among the fastest community detection algorithms. 
Our framework provides a new perspective on community detection methods. It only relies on the notion of local dominance, which is identified solely from local information from the topology. 
It does not need to iteratively optimize an objective function \cite{blondel2008fast,rosvall2008maps,peel2017ground,duch2005community,mclachlan2007algorithm} based on a global randomized null model \cite{newman2006modularity,blondel2008fast,peel2017ground} or resorting to iterative spreading dynamics \cite{raghavan2007near,frey2007clustering} as other state-of-the-art algorithms. 
It is important to emphasise that the communities that are uncovered by LS are not necessarily associated to a high density of links, as in modularity optimisation, or specific patterns of connectivity inside versus across groups, as in methods based on stochastic block models \cite{holland1983stochastic,newman2016structure,schaub2020hierarchical}, but are instead obtained as a group of nodes that are dominated by the same leader.  


\subsection*{Data availability}
All network and vector datasets needed to evaluate the conclusions in the paper are publicly available and present in the paper and/or the Supplementary Materials. The original cellphone datasets of Dakar and Abidjan are accessed through the D4D challenge, and the Beijing dataset is obtained from a Chinese telecommunication operator and the original dataset is not publicly available. Computer code for the LS algorithm and code implementing the analysis described in this paper and other information will be online at \url{https://github.com/UrbanNet-Lab/LS_for_CommunityDetection_and_Clustering}.  


\subsection*{Acknowledgements}
R.Li acknowledges helpful discussions with Dr. Gezhi Xiu and Dr. Wenyi Fang from Peking University. F.S. acknowledges technical help from Mr. Ankang Luo, and Ms. Chenxin Liu from UrbanNet Lab. 
This work receives financial supports from the National Natural Science Foundation of China (Grant No. 72371014, 61903020). L.L. acknowledges the financial supports from the XPLORER PRIZE, and the Special Project for the Central Guidance on Local Science and Technology Development of Sichuan Province (Grant No. 2021ZYD0029). R.L. acknowledges support from the EPSRC Grants EP/V013068/1 and EP/V03474X/1. 
\subsection*{Author contributions}
R.Li conceived the research. R.Li, B.C., H.E.S., L.L., P.E., T.E, and R.L. designed the research. R.Li and F.S. designed the first version of the LS algorithm, T.E. further refined and improved the algorithm. F.S. and D.S. implemented the LS algorithm and conducted experiments. R.Li, F.S., D.S., B.C., P.E., L.L., T.E. and R.L. discussed the results and wrote the manuscript. R.Li was the lead writer of the manuscript. All authors reviewed the manuscript. 

\subsection*{Additional information}
\textbf{Supplementary Information} accompanies this paper at \\
\noindent
\textbf{Competing interests: } The authors declare that they have no competing interests.

\end{document}